\documentclass[preprint]{vgtc}               
\ifpdf
  \pdfoutput=1\relax                   
  \usepackage{graphicx}                
  \DeclareGraphicsExtensions{.pdf,.png,.jpg,.jpeg} 
\else
  \usepackage{graphicx}                
  \DeclareGraphicsExtensions{.eps}     
\fi%

\graphicspath{{figures/}{pictures/}{images/}{./}} 

\usepackage{float}                 
\usepackage{microtype}                 
\PassOptionsToPackage{warn}{textcomp}  
\usepackage{textcomp}                  
\usepackage{mathptmx}                  
\usepackage{times}                     
\usepackage{cite}                      
\usepackage{tabu}                      
\usepackage{booktabs}                  

\usepackage{csquotes}                  
\usepackage{todonotes}
\usepackage{xspace}
\usepackage[normalem]{ulem}

\newcommand{\etal}{\emph{et al.}\@\xspace}
\newcommand{\ie}{\emph{i.e.}\xspace}
\newcommand{\eg}{\emph{e.g.}\xspace}

\onlineid{1141}

\vgtccategory{Research}
\vgtcpapertype{Evaluation}

\vgtcinsertpkg

\title{VisQuiz: Exploring Feedback Mechanisms to Improve Graphical Perception}

\author{Ryan Birchfield 
\and Maddison Caten  %
\and Errica Cheng  %
\and Madyson Kelly%
\and Truman Larson%
\and Hoan Phan Pham%
\and Yiren Ding  %
\and No\"{e}lle Rakotondravony %
\and Lane Harrison %
}
\affiliation{
\scriptsize 
Worcester Polytechnic Institute\thanks{emails: \{rhbirchfield, mrcaten, eycheng, mpkelly, tlarson, ppham, yding5, ntrakotondravony, ltharrison\}@wpi.edu} 
}

\shortauthortitle{Birchfield \MakeLowercase{\textit{et al.}}: VisQuiz: Exploring Feedback Mechanisms to Improve Graphical Perception}

\teaser{
  \centering
  \includegraphics[width=0.8\linewidth]{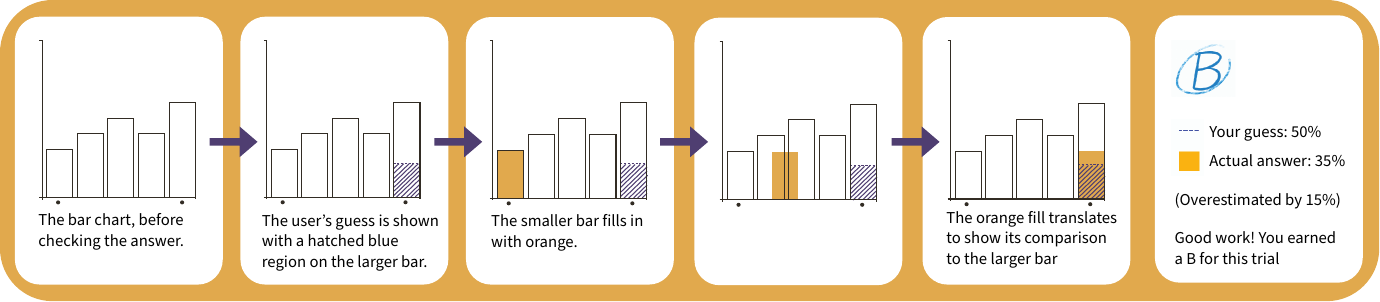}
  \caption{
  In VisQuiz, we explore feedback for graphical perception tasks. Animation and visual design were found to be crucial elements in guiding user attention to reflect on their performance across trials. Similar animation feedback sequences were designed for pie, stacked bar, and bubble charts.
  }
  \label{fig:teaser}
}

\abstract{
In this paper, we explore the design and evaluation of feedback for graphical perception tasks, called VisQuiz.
Using a quiz-like metaphor, we design feedback for a typical visualization comparison experiment, showing participants their answer alongside the correct answer in an animated sequence in each trial, as well as summary feedback at the end of trial sections.
To evaluate VisQuiz, we conduct a between-subjects experiment, including three stages of 40 trials each with a control condition that included only summary feedback. 
Results from $n=80$ participants show that once participants started receiving trial feedback (Stage 2) they performed significantly better with bubble charts than those in the control condition. This effect carried over when feedback was removed (Stage 3).
Results also suggest an overall trend of improved performance due to feedback. 
We discuss these findings in the context of other visualization literacy efforts, and possible future work at the intersection of visualization, feedback, and learning.
Experiment data and analysis scripts are available at the following repository \url{https://osf.io/jys5d/}
}  

\CCScatlist{ 
\CCScatTwelve{Visualization}{Graphical Perception}{Feedback}{}
}

\begin{document}


\firstsection{Introduction}

\maketitle


Visualization studies often use low-level graphical judgement tasks to evaluate competing techniques and to explore perceptual and cognitive facets of visualization.
Graphical perception experiments have compared visualization techniques \cite{cleveland1984graphical, heer2010crowdsourcing}, explored chart reading strategies \cite{kosara2019evidence}, and behavioral factors like social information \cite{hullman2011impact}. 
Other visualization studies have examined visual channels \cite{mccoleman2021rethinking}, network reading tasks \cite{nobre2020evaluating}, and judgments with icon arrays \cite{xiong@2020biased}.
What these studies have in common are that they involve low-level, trial based experiment designs in which participants make multiple judgements about a given visualization.

Another common factor in these studies are short training sessions, often designed to ensure task understanding and to help participants reflect on performance prior to the full study.
However, visualization studies typically do not test the extent to which people might be able to improve their performance in training sessions.
If people can improve visualization task performance through training, there may be implications for empirical visualizations studies--- \eg reducing participant variance before comparing visualization techniques, and visualization literacy efforts--- \eg helping people understand when and how they make mistakes in reading visualizations.
Exploring this dynamic requires more systematic investigation of feedback for visualization tasks.

In this paper, we explore and evaluate feedback mechanisms for a typical graphical perception experiment.
Beginning with the often-replicated comparison tasks from Cleveland \& McGill \cite{cleveland1984graphical}, we augment a 3-stage, 120-trial version of this experiment to include two types of feedback.
In Stage 2, participants receive animated feedback and ``grades'' for each of their trials (\autoref{fig:trial-round-2-feedback}).
At the end of Stage 1, 2, and 3, participants receive an interactive ``report'' showing grades on each of their trials (Figure \autoref{fig:round-feedback}).
\textit{VisQuiz} aims to provide participants with opportunities to improve their visualization performance.

\begin{figure*}[ht]
    \centering
    \includegraphics[width=0.75\textwidth]{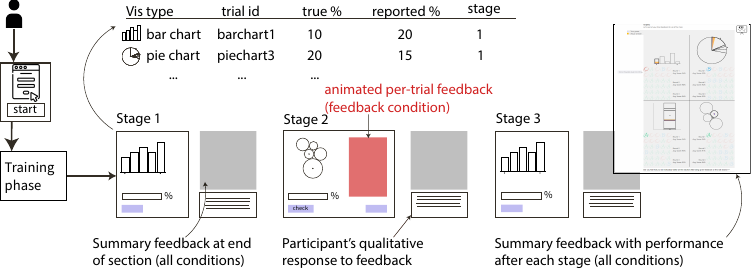}
    \caption{A multistage quiz-style experiment: each stage consists of 40 visualization comparison tasks followed by a feedback page displaying the participant's accuracy score for each type of visualization, its associated letter grade, an animated replay of how each trial was answered, and a space for participants to type how they felt about their performance. In Stage 2, in the feedback condition, participants additionally receive {\color[HTML]{d12222}{animated per-trial feedback}} after each answer. For each trial, the type of visualization, the true percentage, the participant's answer are recorded. }
    \label{fig:methodology}
\end{figure*}

Results of $n=80$ participants in a between-subjects design show no differences in Stage 1 (before feedback was applied), but significant improvements in Stage 2 (per-trial feedback) and Stage 3 (no feedback) in the bubble chart condition, when compared to the control.
Across other chart conditions (bar, pie, stacked bar), results suggest a general trend of improved performance.
Qualitative feedback from participants suggest that participants may actively reflect on their performance and adopt different strategies when given feedback. 
We discuss how these findings may hold implications for ongoing visualization literacy efforts--- particularly those in educational settings--- and how similar feedback schemes may help reduce participant noise in traditional visualization experiments.

\section{Related Work}

Building on crowdsourced studies from Heer and Bostock \cite{heer2010crowdsourcing}, Kong \etal investigate the effectiveness of different treemap visualizations \cite{kong2010perceptual}. Similarly, Talbot \etal investigate bar charts and why people make errors in comparison judgments \cite{talbot2014four}.
Given the common stimuli styles and methodology, we adopt the general experiment design of these studies in our investigation of the impact of feedback.
Other similar visualization studies with low-level tasks could be feasibly reframed in the context of feedback and training. 
For example, Bayesian reasoning tasks common in medicine have been found to be notoriously difficult by both Micallef \etal  \cite{micallef2012assessing} and Ottley \etal \cite{ottley2015improving}, and any potential improvements to these could have tangible benefits for medical decision making.
Other studies involve various aspects of risk and uncertainty (\eg \cite{franconeri2021science, fernandes2018uncertainty}), which could be treated similarly.

The growing area of efforts in visualization literacy also informs our investigation of feedback (for a review, see Firat \etal \cite{firat2021interactive}).
Many visualization literacy approaches focus on assessment, such as the Visualization Literacy Assessment Test from Lee \etal \cite{lee2016vlat} and similarly focused test from Boy \etal \cite{boy2014principled}.
Other efforts focus on interventions with target populations, such as game promoting visualization literacy \cite{alper2017visualization}, classroom activities \cite{koedinger2001toward}, and training for advanced visualization tools \cite{denisova2020treemap}.
In contrast to these studies, which focus on high-level measures of learning, we seek to blend the quantitative nature of empirical visualization studies with literacy efforts, and in doing so investigate more granular means for assessing improvements in visualization performance, similar to question-sets and test scores common in education.

\section{Methodology and Design}

We begin by extending the graphical comparison task from Cleveland and McGill \cite{cleveland1984graphical} and replications (\eg \cite{talbot2014four, heer2010crowdsourcing}), making necessary changes to evaluate the impact of feedback on peoples' performance.
We detail major changes needed, including the design of various types of feedback, a multi-stage intervention style design, and control condition.
Like previous studies, two elements in the tested charts are marked, and participants are asked to estimate \emph{``what percentage is the smaller of the larger''}. 
Performance is measured as error--- the absolute difference between the correct answer and the value reported by participants--- which yields a range of possible performance levels.
The fact that people must use estimation may leave room for improvement, particularly if participants find their answering strategy does not align with the given (correct) feedback.

\subsection{Experiment Stimuli}
We use four chart types in this experiment: stacked bar, bar, bubble, and pie.
The chosen chart types have desirable features for evaluating feedback. 
Talbot \etal found several possible sources of error for bar charts and stacked bar charts, like the distance between bars being compared \cite{talbot2014four}.
Kosara provides evidence that people may use different visual cues when reading pie charts, another source of error and possible room for improvement \cite{kosara2019evidence}.
Finally, multiple studies of bubble chart encodings in the mapping domain have found that people tend to underestimate area, a systematic bias that might be mitigated with feedback \cite{montello2002cognitive}.

Each chart shows five data points and participants are asked to compare two elements, randomly chosen and marked with dots. 
Results from Talbot \etal indicate that people's estimated percentages will typically consist of multiples of 5 or 10 \cite{talbot2014four}. 
Given this, we adopt a data generation strategy that spans a range of answers from 5-95.
We also inform users that all the answers should be multiples of five.
Each chart type appears ten times in a stage, encoding ratio between 5\% and 95\%.
To facilitate comparison between groups, we pre-generate these datasets, and randomize trial order throughout each stage.
All participants receive the same datasets (randomized), facilitating comparison between participants.
Each trial results in several outputs, including the type of chart, a trial ID number, the true ratio, and the ratio reported by the participant. 

\subsection{Experiment Procedure}

The experiment procedure consists of four phases: an initial training phase, followed by three stages which vary between the feedback and control conditions (see \autoref{fig:methodology}).
In the training phase, participants are given one trial of visualization type to familiarize them with the trial structure, with no feedback in either condition. 
There are 40 trials each in the three main stages of the experiment. 

\begin{figure}[H]
    \centering
    \includegraphics[width=0.9\columnwidth]{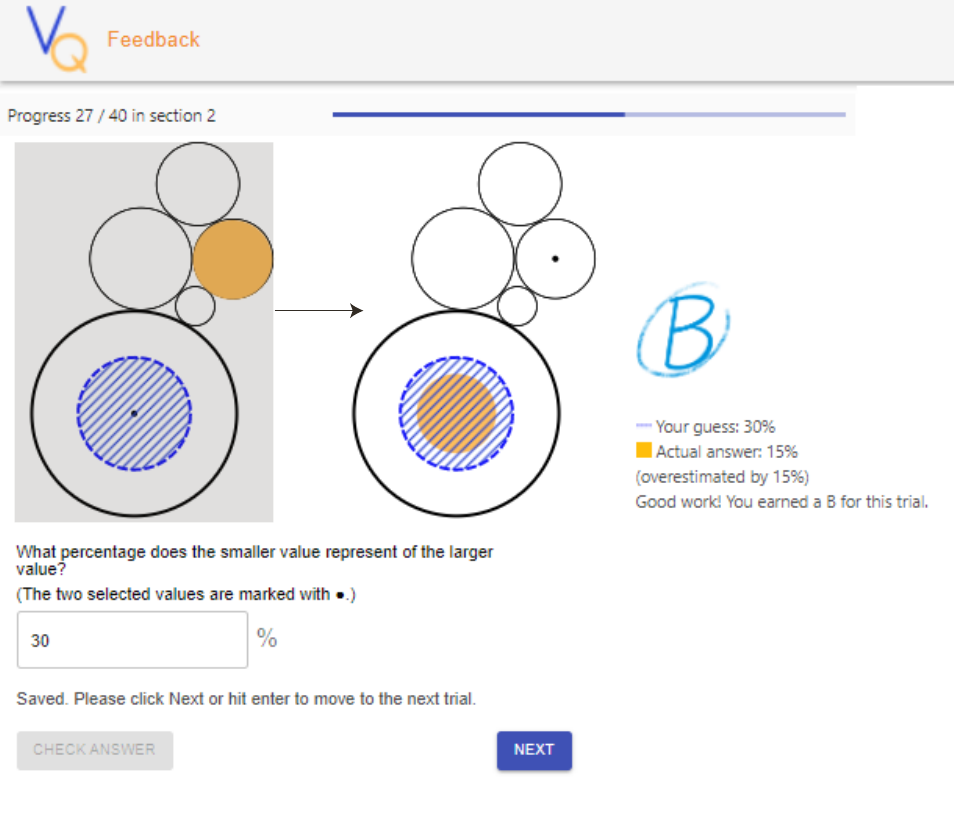}
    \caption{Example task from Stage 2 in the feedback condition showing feedback given after an answer is submitted. The greyed area illustrates a frame of the animation before the final result (right) where the orange circle transitions onto the blue one, contrasting the participant's answer with the correct answer.}
    \label{fig:trial-round-2-feedback}
\end{figure}

\textbf{Stage 1-- Calibration.} 
The calibration stage ensures that, even in the feedback condition, participants will have received 40 trials (10 per chart type) before receiving any kind of feedback on their performance.
This serves a baseline for comparison.
\textbf{Stage 2-- Feedback Enabled/Disabled.}
In the second set of 40 trials, 
participants are either given animated feedback after each trial (feedback condition) or no feedback (control condition). The design of this feedback is detailed in \autoref{sec:designing-feedback} (also see \autoref{fig:trial-round-2-feedback}). 
\textbf{Stage 3-- Quiz.}
In the final stage, participants again answer questions without any in-trial feedback. 
Having a third stage, sans feedback, provides a way to evaluate whether any possible effect of feedback is carried over from the previous stage. 
Each stage ends with a summary feedback page that details how the participant performed on each trial of the current stage (and previous stages, if applicable). 
Participants are also encouraged to take a break between each stage, if needed. 
\textbf{Feedback Perceptions \& Demographics.}
At the end of the experiment, participants were asked to reflect on the quality of their feedback and whether they felt it helped them improve their performance--- we use these comments to augment the primary quantitative results.
Participants were also asked to provide demographic information, including reported gender, age, country, and highest degree obtained. 

\subsection{Designing Feedback for Graphical Perception}\label{sec:designing-feedback}

To provide participants with graphical perception feedback, we frame the experiment as a quiz, called VisQuiz.
The quiz format provided a natural framing for providing summary (end of stage) and per-trial feedback (feedback condition).
The overall design challenge for feedback was to provide participants with an effective means to reflect on their performance in the context of specific graphical comparison trials.
Through an iterative design process, we developed feedback with both textual and visual encodings.

\begin{figure}[H]
    \centering
    \includegraphics[width=0.9\columnwidth]{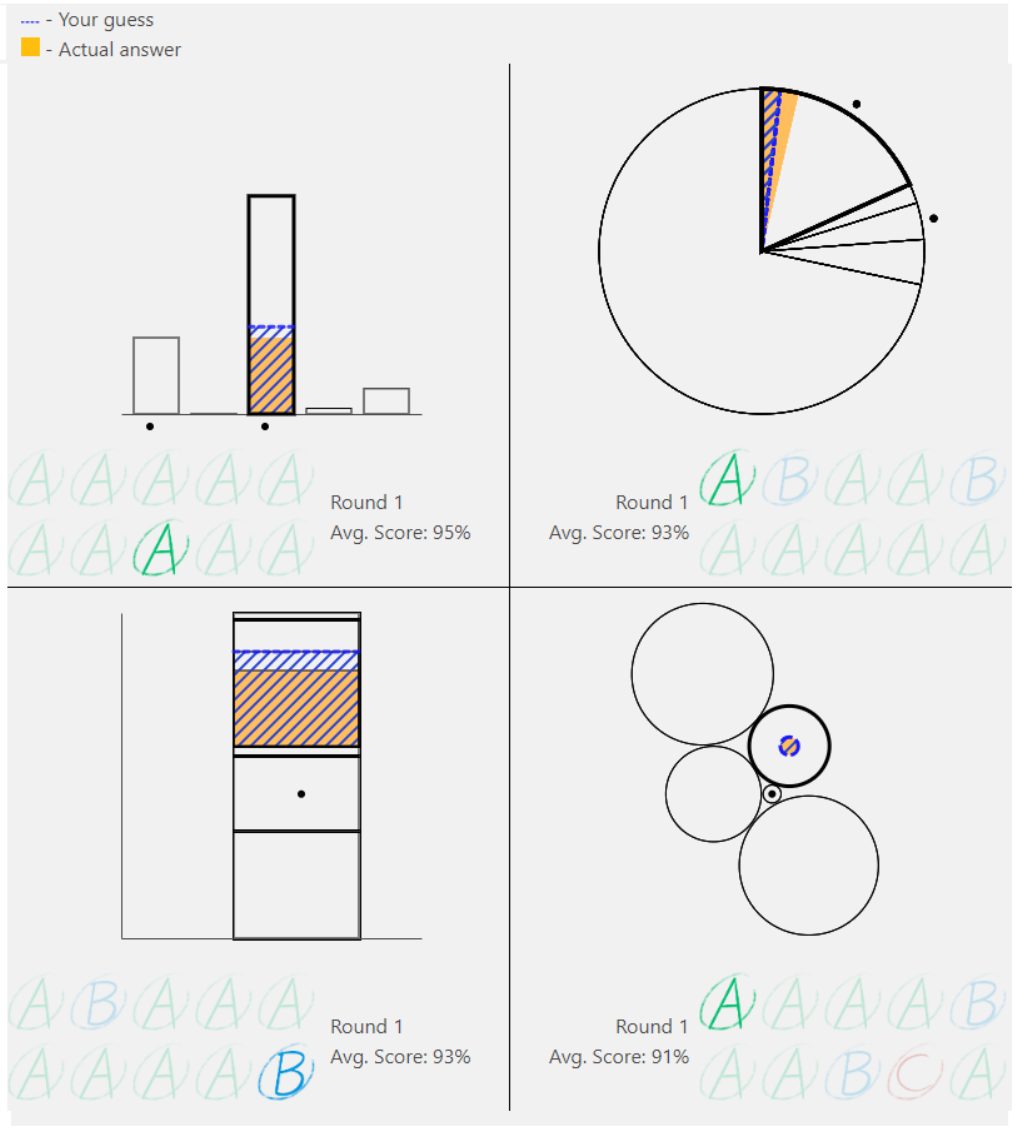}
    \caption{
    Participants receive summary pages showing their performance across all trials in previous sections.
    In addition to a visual summary of trial performance, participants are guided to click on particular trials to replay their answers and animated feedback for further reflection.
    }
    \label{fig:round-feedback}
\end{figure}

Text feedback per trial includes participants' guess (\eg 15\%), the correct answer (\eg 25\%), the amount participants over- or under-estimated (\eg 10\%), and a grade with an affirmation.
Grades and affirmations were inspired by the popular game Guess the Correlation, where players estimate correlation given scatterplots, and are given various amounts of points based on their distance from the true answer \cite{guessTheCorrelation}.
Grades were A (0-5\% error), B (10-15\%), and C (15\%+), with corresponding affirmations such as ``Good work!'' and ``Missed the mark\ldots''.
Pilot studies demonstrated that textual feedback alone was perceived as helpful, yet insufficient to help participants think of new strategies. 
A visual approach was needed.

\subsubsection{Animated Per-Trial Feedback Design}\label{subsec:in-trial-feedback}

To facilitate reflection during feedback-enabled trials, we design an animated feedback sequence, triggered after each participant answer.
Upon answering, participants see the textual feedback alongside an animation relating the two elements compared in the context of the participant's guess and the true answer (see \autoref{fig:trial-round-2-feedback}).
The animation sequence first superimposes the participant's guess on the target element using a blue hatched design. 
The texture distinguishes the participant guess from actual elements of the chart, while providing partial transparency for layering.
After a pause to allow the participant to see their guess in the context of the target, the smaller element is filled (orange), and animated from its original position to align with the larger element.
The end state provides participants with both visual and textual representations of any over- or under-estimation they made.
In addition to guiding user attention through sequencing, all animated transitions align with the type of chart used: pie slices rotate around the center, stacked bar elements move along the y-axis, bars along the x-axis, and bubbles move directly through x-y space to their target.

\subsubsection{Summary Feedback Design}

We provide a summary feedback view at the end of each stage in both conditions.
This design was partially inspired by studies on LabInTheWild, which uses summary feedback views on various experiments to motivate volunteers to participate in studies \cite{reinecke2014quantifying}.
Shown in \autoref{fig:round-feedback}, the summary feedback display consists of a section per chart type, and uses grade glyphs to represent each trial. Overall accuracy is represented as a percentage score, with 100\% being perfect (0\% was found to be confusing during testing, so the score was inverted). 
Interaction was key in the summary feedback view, as hovering over a grade would replay the animation sequence, showing both the trial and participants' answers.
Taken together, per-trial and summary feedback provided participants with multiple opportunities to improve their performance between trials and/or stages.

\begin{figure*}
    \centering
    \includegraphics[width=0.9\textwidth]{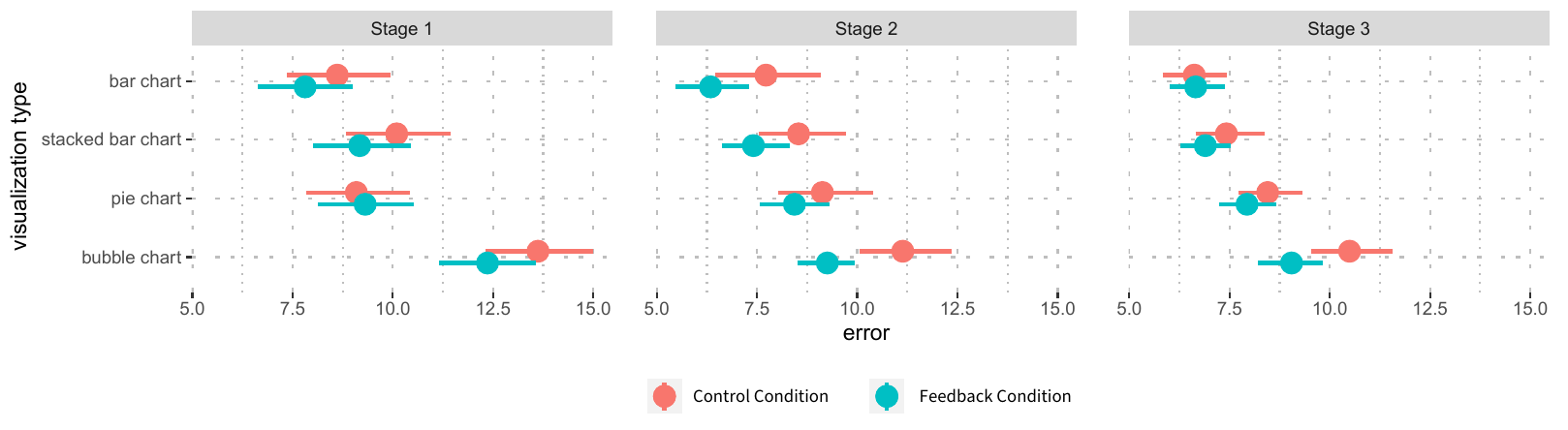}
    \caption{
    Results of the experiment comparing feedback condition $n=40$ to control condition $n=40$ across three stages of 40 trials each. Error in Stage 1 is largely similar, with differences emerging between groups as feedback is enabled (Stage 2) and then removed again (Stage 3).
    }
    \label{fig:error-in-chart}
\end{figure*}

\section{Results}

Participants were recruited through the Prolific.co platform.
Pilot testing revealed that participants took on average about 30 minutes to complete the study.
Based on this estimate, we paid \$7.40 per participant.
Some participants did not complete the study, and were replaced. 
In total, we recruited 40 participants for the control group, and 40 participants for the feedback-enabled group.

Each participant answered 120 trials across three stages in the experiment. 
Collected data include the type of visualization, stage ID (1, 2, or 3), the correct answer, and the participant's guess. 
Error is calculated as $error = \vert truePercentage - guessedPercentage \vert $. 
Approximately 6\% of responses were not multiples of 5. Since these were not obviously negligent behavior-- many were close to the correct answer-- we include them in the analysis.

\subsection{Quantitative Analysis} In \autoref{fig:error-in-chart}, we observe lower error in the feedback condition, compared to the control condition.
Errors are similar in Stage 1.
However, in Stages 2 and 3, there appear to be more differences between the control and feedback conditions.
To evaluate the effect of feedback on participants' performance across different charts, \ie whether the feedback leads to lower error, we conduct a between-group analysis comparing the control (summary only) group with the feedback (per-trial feedback) group. 
Student t-test comparisons between participants' performance across conditions show a significant effect for \emph{bubble charts} during Stage 2 ($mean_{control} = 11.13$, $mean_{treatment} = 9.25$, $p = 0.0046$, $df = 684.42$),
and Stage 3 ($mean_{control} = 10.49$, $mean_{treatment} = 9.04$, $p = 0.013$, $df = 766.71$).
An effect is also found for \emph{bar charts} in Stage 2 ($mean_{control} = 7.72, mean_{treatment} = 6.34, p = 0.04, df = 709.79$).
In all cases, the feedback condition outperforms the control condition (\ie the feedback condition shows lower error).

To analyse the extent to which possible effects of feedback are carried between stages, we conduct an analysis of participant's error while progressing across the three stages in VisQuiz. 
This is a within-subjects analysis for the treatment group, across stages.
A one way ANOVA revealed significant differences in the error across the three stages for stacked bar chart ($F = 6.30, df = 2, p = 1.9 E-3$) and bubble chart ($F = 13.76, df = 2, p = 1.23 E-6$).
Post-hoc pairwise comparisons show that error is significantly lower for the stacked bar chart between Stage 1 and 2 ($est = -1.76, 95\% CI = [-3.34, -0.18], p.adj = 2.42E-2$), 
and Stage 1 and 3 ($est = -2.28, 95\% CI = [-3.86, -0.69], p.adj = 2.12E-3$). Similarly, error is significantly lower for bubble charts between 
Stage 1 and 2 ($est = -3.11, 95\% CI = [-4.77, -1.44], p.adj = 3.67E-5$), 
and Stage 1 and 3 ($est = -3.31, 96\% CI = [-4.97, -1.65], p.adj = 9.41E-6$).

\subsection{Qualitative Analysis}  In the feedback condition, participants were also asked to (optionally) reflect on the quality of the feedback and how they used it.
Curiously, out of 15 participants who left comments, 5 mentioned bubble charts specifically, far more than any other chart in the provided comments. For example, one participant notes a switch between radius and area encodings:

\vspace{-0.5em}

\begin{displayquote}
    \emph{
As far as the circles went, i was confused with the objective, i was between radius and total area, so in the beginning i went with area but then switched to radius as my basis, so in that test i did worse than i expected, on the other, i performed as i had expected    
}
\end{displayquote}

\vspace{-0.5em}

Other participants left general positive remarks about the presence of feedback, for example:
\vspace{-0.5em}

\begin{displayquote}
    \emph{
Honestly I felt better when receiving live feedback. It definitely felt harder this last stage losing the ability to adjust according to the last figure's feedback.
}

    \emph{
Yes, learning from the feedback given is very helpful in the process of rearranging your mind and the way you look at these graphs
}
\end{displayquote}
\vspace{-0.5em}

Several participants preferred to contextualize their performance in terms of the grades, \ie the count of As, Bs, and Cs, they received:

\vspace{-0.5em}

\begin{displayquote}
    \emph{
I think I did better in roughly accessing the percentages in a faster way, shifting one shape into the other in my imagination. But I also made more obvious mistakes (the Cs with very large margins of error) that I didn't notice until I got to the feedback, probably once again due to my concentration decreasing due to the length of the activity.
}
\end{displayquote}

\vspace{-0.5em}

Finally, other participants left suggestions on how to further improve usability, for example playing feedback animation on a loop.

\section{Discussion and Conclusion}

Results of the VisQuiz evaluation suggest that interactive, animated feedback can have a positive impact on peoples' graphical perception performance.
Participants generally improve in both conditions, presumably due to the provided trial and summary feedback. Greater improvements are seen with per-trial feedback, however.

Visualization performance feedback may not benefit all equally, or universally. This experiment focuses on groups, which raises questions about the extent to which an \emph{individual} might be able to improve their performance.
Hierarchical modeling is one possible approach to investigating such effects, which has been successfuly used in other visualization studies \cite{kay@2016beyond, fernandes2018uncertainty}.
Feedback effectiveness may also vary between chart types. We observe more improvement in bubble charts, perhaps due to systematic underestimation identified in prior work \cite{montello2002cognitive}. Other chart types require more investigation. Pie charts, for example, may be interpreted using several different methods \cite{kosara2019evidence}, which might suggest more extensive training or even adaptive training is needed.

Target audience is another important consideration. Here, we test on crowdsourced participants, who may have varied motivation for improving their ability to read charts.
However, as visualization learning activities become a deeper part of schooling education (\eg \cite{firat2021interactive, panavas2022juvenile}), animated feedback and quantitative assessments of performance could become an important tool for promoting and training visualization literacy. Studies might also look short- versus long- term gains, similar to Bateman \etal's studies on visualization memorability \cite{bateman2010useful}, and recent advances in visualization onboarding techniques \cite{stoiber2021design}.

Similar animated feedback could be developed for other low-level tasks, such as correlation comparison \cite{yang2018correlation}, or for more critical higher-level tasks like medical decision-making involving risk \cite{micallef2012assessing, ottley2015improving}. 
The possibility of not only measuring the relative effectiveness of visualization techniques, but improving peoples' ability to use them, might be a general model applicable across many visualization studies.

\acknowledgments{
    This project was supported by NSF grant $\#1815587$.
}

\bibliographystyle{abbrv-doi}

\bibliography{paper}
\end{document}